\documentclass{article}   

\usepackage{amsmath, amssymb, graphics, graphicx}
\usepackage{textcomp, subfigure}
\usepackage{float}     

\usepackage{array}
\usepackage[strict]{changepage} 
\usepackage[graphicx]{realboxes}
\usepackage{placeins}
\usepackage{lscape}
\usepackage{lineno}    

\usepackage{graphicx}  
\usepackage{dcolumn}   
\usepackage{bm}        
\usepackage{amssymb}   

\usepackage{amsmath}   
\usepackage{paralist}  

\usepackage[numbers, sort&compress]{natbib}

\usepackage{color, soul}

\usepackage[labelfont=bf]{caption} 
\usepackage[strict]{changepage} 
\usepackage{textgreek} 

\usepackage{lscape}

\usepackage[strict]{changepage}
\usepackage[labeled]{multibib}
\newcites{S}{Supplementary References}
\usepackage{titling}

\newcommand{\um}{{\textmu}m }
\newcommand{\ket}[1]{\ensuremath{\left|#1\right\rangle}}

\newcommand{\kex}{\ensuremath{\kappa_{\textrm{ex}}}}
\newcommand{\sinf}[1]{\ensuremath{s_{#1,\textrm{in}}}}
\newcommand{\soutf}[1]{\ensuremath{s_{#1,\textrm{out}}}}

\usepackage[table]{xcolor}
\usepackage{makecell}
\usepackage{boldline}
\setcellgapes{3pt}

\begin{document}
	
	\title{Complete linear optical isolation at the microscale with ultralow loss}

	\author{JunHwan Kim$^\dagger$, Seunghwi Kim$^\dagger$, Gaurav Bahl$^\ast$\\
		\\
		\footnotesize{Mechanical Science and Engineering,}\\
		\footnotesize{University of Illinois at Urbana-Champaign, Urbana, Illinois, USA}\\
		\footnotesize{$^\dagger$ Equal contribution} \\
		\footnotesize{$^\ast$ To whom correspondence should be addressed; E-mail: bahl@illinois.edu.}
	}
	
	\date{}
	\maketitle
		
		\begin{abstract}
			Low-loss optical isolators and circulators are critical nonreciprocal components for signal routing and protection, but their chip-scale integration is not yet practical using standard photonics foundry processes. 
			%
			The significant challenges that confront integration of magneto-optic nonreciprocal systems \cite{Wolfe1990, Ballato1995} on chip \cite{Tien2011,Bi2011,Shoji2008,Goto2012} have made imperative the exploration of magnet free alternatives \cite{Yu2009, 5784283, Poulton2012, Fang:2012dx, Zhu:2013jf, Sounas:2013ey, Sounas2014, Kang:2011el, Lira:2012ck, Fan:2012go, Peng:2014kl,Tzuang:2014hu, Sayrin:2015bm}. 
			However, none of these approaches have yet demonstrated linear optical isolation with ideal characteristics over a microscale footprint -- simultaneously incorporating large contrast with ultralow forward loss -- having fundamental compatibility with photonic integration in standard waveguide materials.
			Here we demonstrate that complete linear optical isolation can be obtained within any dielectric waveguide using only a whispering-gallery microresonator pumped by a single-frequency laser. 
			The isolation originates from a nonreciprocal induced transparency \cite{Kim2015} based on a coherent light-sound interaction, with the coupling originating from the traveling-wave Brillouin scattering interaction, that breaks time-reversal symmetry \cite{Jalas2013} within the waveguide-resonator system.
			Our result demonstrates that material-agnostic and wavelength-agnostic optical isolation is far more accessible for chip-scale photonics than previously thought.
		\end{abstract}

		
		Ideal optical isolators should exhibit complete linear isolation -- {where completeness implies} perfect transmission one way (i.e. zero forward insertion loss) and zero transmission in the opposite direction -- without any mode shifts, frequency shifts, or dependence on input signal power. 
		Unfortunately, the established magneto-optical techniques \cite{Wolfe1990, Ballato1995} for achieving nonreciprocal optical transmission \cite{Jalas2013} have proven challenging to implement in chip-scale photonics due to fabrication complexity, difficulty in locally confining magnetic fields, and material losses \cite{Tien2011,Bi2011,Shoji2008,Goto2012}. 
		In light of this challenge, several non-magnetic alternatives for breaking time-reversal symmetry \cite{Jalas2013} have been explored both theoretically \cite{Yu2009, Poulton2012, Fang:2012dx, Zhu:2013jf, Sounas:2013ey, Sounas2014} and experimentally \cite{Kang:2011el, Lira:2012ck, Fan:2012go, Tzuang:2014hu, Peng:2014kl, Sayrin:2015bm}.
		However, various limitations still persist with the proposals that are compatible with chip-scale fabrication. 
		Nonlinearity-based nonreciprocal isolators are fundamentally dependent on input field strength \cite{Fan:2012go, Peng:2014kl} and hence do not generate linear isolation \cite{Shi:2015cu}.
		Dynamic modulation \cite{Yu2009,Fang:2012dx} is a powerful approach that generates linear isolation, but current chip-scale demonstrations are still constrained by extremely large forward insertion loss and low contrast \cite{Lira:2012ck,Tzuang:2014hu}. 
		Finally, the Brillouin acousto-optic scattering approach \cite{5784283, Poulton2012} {is based on using stimulated intermodal scattering enabled by phonons to induce unidirectional photonic loss. This technique} promises linearity and easy implementation in most optical materials but requires waveguide lengths of several centimeters {to several meters} \cite{Kang:2011el} for practical isolation. Recent advancements in on-chip gain \cite{VanLaer:2015wg, Kittlaus:2016go} {have improved the future potential of miniaturizing this approach, but till date no microscale SBS isolator has been demonstrated.}
		{Presently, there is no experimentally demonstrated magnet-free technique that can provide linear optical isolation, with ultra-low loss, and a microscale footprint at the same time.}
		{Comparisons against state-of-the-art experimental results on non-magnetic microscale isolation can be found in Table S.1 of the Supplement.}

		In this work we explore a fundamentally different approach {based on an induced transparency process, generated by destructive optical interference enabled by a non-radiative traveling-wave acoustic coherence within a resonator-waveguide system} \cite{Kim2015, Dong2015}, {which we describe below}. 
		This phenomenon is termed Brillouin scattering induced transparency (BSIT) \cite{Kim2015} and is an acousto-optic analogue of electromagnetically induced transparency (EIT) {in which the high coherence electronic state is replaced by a traveling phonon mode.} {In BSIT, the} momentum conservation requirement between photons and phonons helps break time-reversal symmetry. 
		Using this concept we demonstrate that complete linear optical isolation can be achieved in a simple photonic microsystem, composed only of a silica waveguide and silica microresonator in the 100 \um regime, and can be dynamically reconfigured on demand.
		The mechanism is available in all dielectrics and can thus, in principle, be implemented with any waveguide material available in photonics foundries.
		We show theoretically that when operating within the strong acousto-optical coupling regime, the BSIT system can enable perfect {lossless} transmission of light in the forward direction in a waveguide, while maintaining  complete absorption in the reverse direction -- {the condition of complete linear isolation}.
		Experimentally, we demonstrate a device {operating very close to the strong coupling regime} and capable of generating an enormous 78.6 dB of isolation contrast per 1 dB of forward insertion loss within the induced transparency bandwidth.

		\begin{figure}[p]
			\begin{adjustwidth}{-1in}{-1in}
				\vspace{-40pt}
				\centering
				\includegraphics[width=1.2\textwidth]{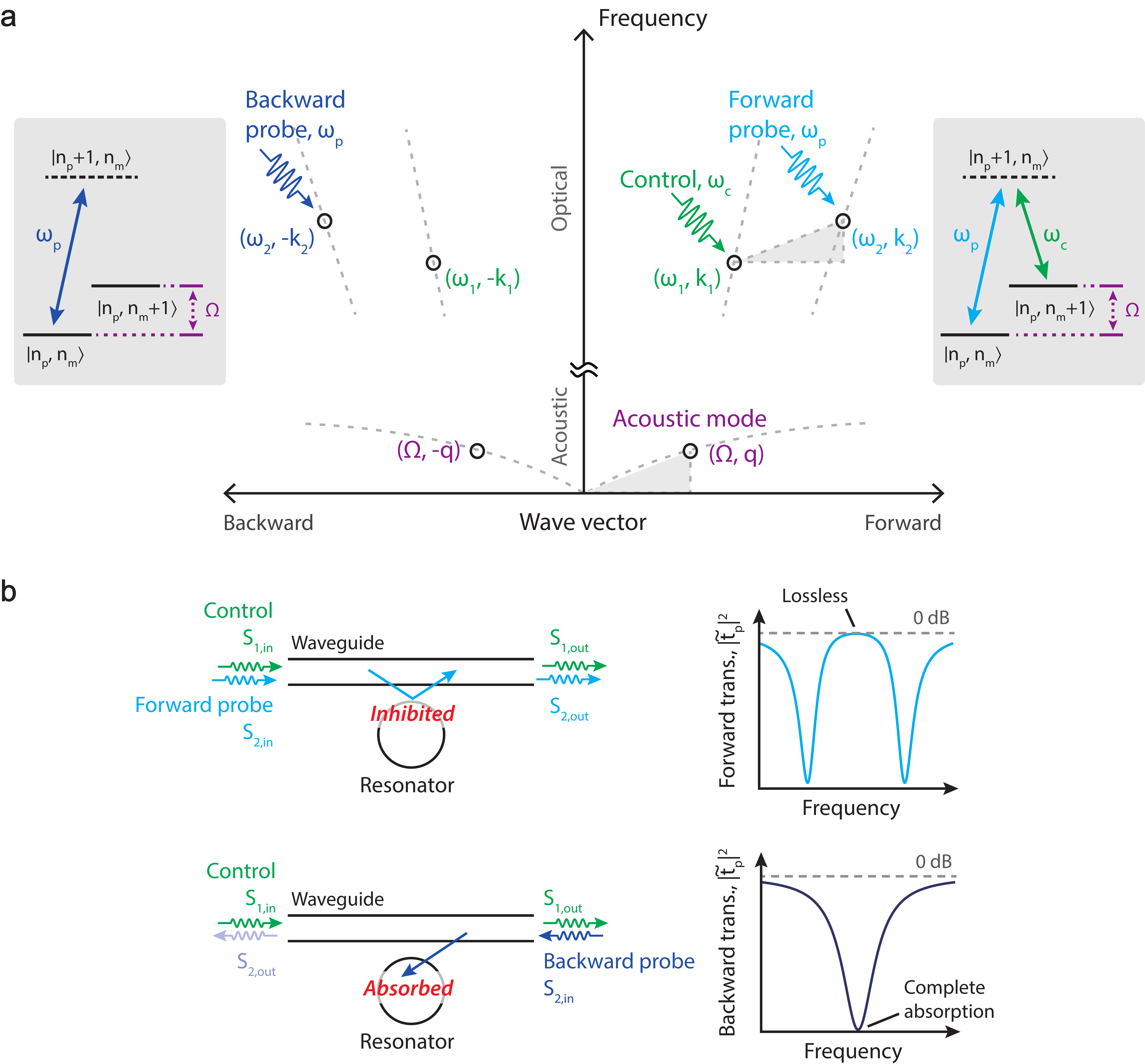}
				\caption{
					{\textbf{Achieving optical isolation through non-reciprocal Brillouin scattering induced transparency in a whispering-gallery resonator: }}
					\textbf{a.} 
					The interference of excitation pathways in the BSIT system are described through an energy-level picture (grey boxes), using probe photon number $n_p$ and phonon number $n_m$. 
					Absorption of a probe photon into the ($\omega_2$, k$_2$) optical resonance is modeled as an effective transition $\ket{n_p,n_m} \rightarrow \ket{n_p+1,n_m}$. In presence of the control field, the probe photon could scatter to the lower resonance ($\omega_1$, k$_1$) while adding a mechanical excitation in ($\Omega$, q), which is an effective transition to state \ket{n_p,n_m+1}. 
					However, the coherent anti-Stokes scattering of the control field from this mechanical excitation would generate an interfering excitation pathway for the original state \ket{n_p+1,n_m}. This process is analogous to EIT and results in a window of transparency for the forward optical probe, inhibiting the original $\ket{n_p,n_m} \rightarrow \ket{n_p+1,n_m}$ absorption transition. 
					The necessary momentum matching requirement, not visible in the energy diagram, is represented using the dispersion relation (middle) to elucidate the breaking of time-reversal symmetry for the probe signal.
					\textbf{b.} We implement this mechanism using a waveguide and a whispering gallery resonator, in which probe signals tuned to either of the ($\omega_2$, $\pm$k$_2$) optical resonances are typically absorbed by the resonator under the critical coupling condition. The presence of a forward control field, however, creates the BSIT interference \cite{Kim2015} only for forward probe signals and inhibits absorption. Under strong acousto-optical coupling, the waveguide-resonator system is rendered lossless at the original resonance.}
				\label{wk_3lvl}
			\end{adjustwidth}
		\end{figure}
		
		\vspace{12pt}
		
		Let us first qualitatively discuss how ideal optical isolation can be achieved by means of the BSIT light-sound interaction in dielectric resonators \cite{Kim2015, Dong2015}.
		We consider a whispering-gallery resonator having two optical modes ($\omega_1$,~k$_1$) and ($\omega_2$,~k$_2$) that are separated in ($\omega$, k) space by the parameters of a high coherence traveling acoustic mode ($\Omega$, q). This is the requisite phase matching relation for BSIT (Fig.~\ref{wk_3lvl}a), indicating that phonons enable coupling of the photon modes through photoelastic scattering.
		{We stress here that the two modes should belong to different mode families of the resonator in order to ensure that scattering to other optical modes from the same phonon population is suppressed.}
		When this system is pumped with a strong `control' field on the lower optical resonance ($\omega_1$, k$_1$), an EIT-like optomechanically induced transparency \cite{Safavi-Naeini2011, Weis:2010ci} appears within the higher optical resonance ($\omega_2$, k$_2$), due to coherent interference originating from the acousto-optical interaction \cite{Kim2015, Dong2015}.  
		{A description of this interference can be presented both classically} \cite{Kim2015} {or through by a quantum mechanical approach} \cite{Dong2015}. {Briefly, one can consider signal or `probe' photons arriving from the waveguide at frequency $\omega_2$ that are on-resonance and being absorbed by the resonator mode ($\omega_2$, k$_2$). When the control field is present in a BSIT phase-matching situation, these probe photons could scatter to ($\omega_1$, k$_1$) causing a mechanical excitation of the system. However, anti-Stokes scattering of the strong control field from this mechanical excitation will generate a phase-coherent optical field that interferes destructively with the original excitation of the mode at ($\omega_2$, k$_2$). The result is a pathway interference that is measured as an induced optical transparency in the waveguide, where no optical or mechanical excitation takes place, and the resonant optical absorption is inhibited} (Fig.~\ref{wk_3lvl}b~top). {The strength of this interference is set by the intensity of the control laser. The phase of the mechanically dark mode is instantaneously set by the phases of the control and probe optical fields, and does not require phase coherence between them.} 
		It is crucial, however, to note that this transparency in BSIT only appears for probe signals co-propagating with the control laser. 
		Probe light in the counter propagating i.e. time-reversed direction, on the other hand, occupies the high frequency optical mode with parameters ($\omega_{2}$, -k$_{2}$). For BSIT to occur in this case, an acoustic mode having parameters ($\Omega$,~$- (\textrm{k}_1 + \textrm{k}_2$)) would be required for compensating the momentum mismatch between the {forward control and backward probe} optical modes. However, since such an acoustic mode is not available in the system, no interaction occurs for the counter-propagating probe and the signal is simply absorbed into the resonator (Fig.~\ref{wk_3lvl}b - bottom).

		\vspace{12pt}
		
		The classical field equations for coupled light and sound in this waveguide-resonator system are presented in the Supplement \S 1. The transmission coefficient $\tilde{t}_p$ of the probe laser field can be derived as:
		
		\begin{align}
		\tilde{t}_p ~ = ~ \frac{\soutf{2}}{\sinf{2}} ~ = ~ 1 - \frac{ \kex }{\left( \kappa_2/2 + j \Delta_2 \right) +G^{2}/\left( \Gamma_B/2 + j \Delta_B \right)}
		\label{eq:AS_transmission}
		\end{align}
		where \sinf{i} and \soutf{i} are the optical driving and output fields in the waveguide (Fig.~\ref{wk_3lvl}b) at the control ($i$=1) and probe ($i$=2) frequencies. $G$ is the pump-enhanced Brillouin coupling rate manipulated by the control optical field $s_{\textrm{1,in}}$ in the waveguide via the relation $G = | s_{\textrm{1,in}} \, \beta \,\sqrt{\kappa_{\textrm{ex}}} / \left( \kappa_1/2 + j \Delta_1 \right) |$. Here $\beta$ is the acousto-optic coupling rate, $\kappa_{i}$ are the loaded optical loss rates, $\Gamma_B$ is the phonon loss rate, and \kex~is the coupling rate between the waveguide and resonator. The loaded optical loss rates are defined as $\kappa_i = \kappa_{i,\textrm{o}} + \kappa_{ex}$ where $\kappa_{i,\textrm{o}}$ is the loss rate intrinsic to the optical mode. The $\Delta_{i}$ parameters are the field detunings, with subscript $B$ indicating the acoustic field. This response matches the system of optomechanically induced transparency (OMIT) \cite{Safavi-Naeini2011, Weis:2010ci}, with the exception that the pump field is also resonant and the coupling rate $\beta$ is dependent on momentum matching. {As we explain later, the pump resonance significantly enhances the maximum coupling rate $G$ achievable in contrast to single-mode OMIT systems.}

		\begin{figure}[b!]
			\begin{adjustwidth}{-1in}{-1in}
				\centering
				\includegraphics[scale=0.5]{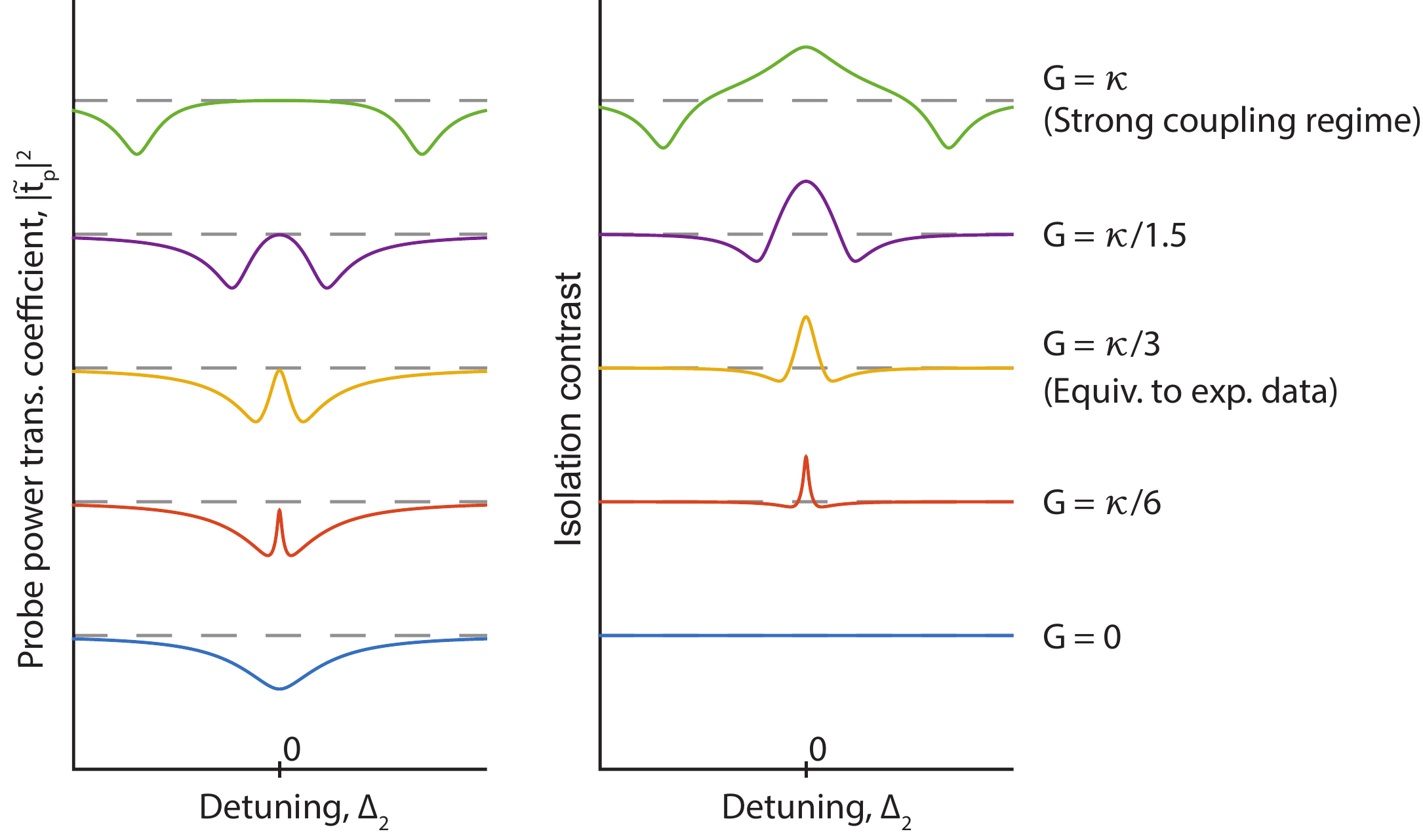}
				\caption{
					\textbf{Evolution of the transparency and isolation contrast as a function of pump-enhanced Brillouin coupling $G$.}
					In the weak coupling regime ($G \ll \kappa$), the transparency linewidth and contrast bandwidth are defined by the acoustic linewidth $\Gamma_{B}$ \cite{Kim2015}. As coupling $G$ increases, the isolation contrast improves, bandwidth is expanded and the optical mode with transparency appears as a splitted mode. In the strong coupling regime, the isolation bandwidth is independent of the acoustic mode and is instead defined by optical mode linewidth $\kappa$ only. The dashed lines indicate the perfect transmission baseline (left) and zero isolation contrast (right) respectively.}
				\label{varyingG}
			\end{adjustwidth}
		\end{figure}

		Equation \ref{eq:AS_transmission} is key to understanding how an ideal optical isolator can be obtained.
		First, we examine the case of no acousto-optic coupling $G=0$, resulting from either modal mismatch ($\beta = 0$) or zero applied control laser power ($\sinf{1} = 0$). 
		In this case Eq.~\ref{eq:AS_transmission} exhibits a well-known Lorentzian shaped transmission dip implying that the probe optical field in the waveguide is simply absorbed by the resonator~\cite{GorodetskyOptimal}. 
		Critical coupling between resonator and waveguide is enabled when $\kex = \kappa_{2,\textrm{o}}$ and results in complete absorption of the probe light from the waveguide at resonance ($\Delta_2 = 0$).
		With critical coupling in place, let us now introduce the effects of the acousto-optic coupling. For very large acousto-optic interaction strength, i.e. $G \rightarrow \infty$, Eq.~\ref{eq:AS_transmission} indicates that we recover perfect transmission $\left| \tilde{t}_p \right|^2 = 1$ even when the waveguide and resonator are critically coupled. Forward propagating probe light in the waveguide, co-propagating with the control laser, can thus transmit perfectly with no absorption at resonance in the ideal case. 
		At the same time, we have no Brillouin coupling ($\beta = 0$) for counter-propagating control and probe optical fields due to the momentum mismatch as indicated previously.
		This implies that, for a counter-propagating probe, the system remains in the critical coupling region resulting in complete absorption.
		Since forward probe signals transmit with zero absorption, and backward probe signals are completely absorbed (Fig.~\ref{wk_3lvl}b), this system is an ideal linear isolator at the transparency resonance.

		A more practically accessible case is $G \ge \kappa_2$, also known as the strong coupling regime~\cite{Groblacher2009}, where the induced transparency grows to the width of the optical mode.
		Strong coupling can be reached for high coherence phonon modes (small $\Gamma_B$) with large acousto-optic coupling $\beta$ and large control driving field \sinf{1}.
		The evolution of the optical transparency and isolation contrast with increasing coupling $G$ is illustrated in Fig.~\ref{varyingG}. In the weak coupling regime ($G \ll \kappa_2$) the isolation contrast is defined roughly by the linewidth of the phonon mode. 
		As $G$ increases the transparency window broadens until eventually reaching the strong coupling regime where the isolation contrast {bandwidth reaches a maximum equaling the optical loss rate $\kappa_2$, as long as the acoustic frequency is higher than this value}. 
		{Thus, the isolation bandwidth can be improved to the several GHz range if a higher frequency acoustic mode is used}~\cite{Bahl:2011cf} {in conjunction with a low-Q (high $\kappa_2$) optical mode, and the reduction in coupling is compensated by other means}~\cite{Dostart2015}.
		In this regime, we also achieve the desired ultra-low forward insertion loss.
		Such large transparency can also be interpreted as the splitting of the optical mode~\cite{Peng2014}.
		The absence or minimization of forward loss necessarily implies linear optical response at frequency $\omega_2$ without any nonlinearity or mode conversion.

		\vspace{12pt}
		
\begin{figure}[t]
			\begin{adjustwidth}{-1in}{-1in}
				\includegraphics[scale=0.3]{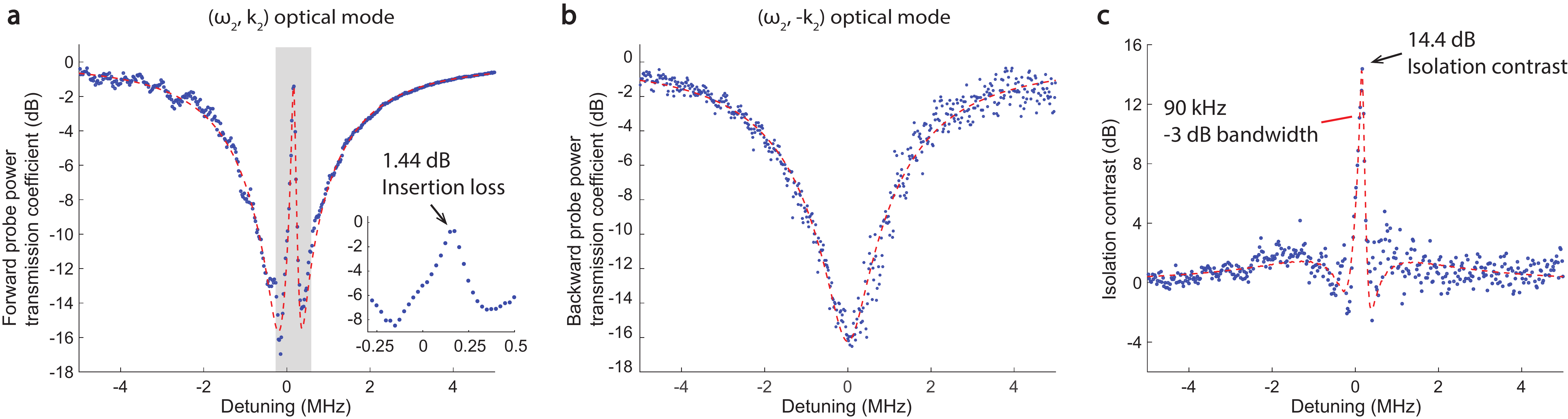}
				%
				%
				\caption{
					\textbf{Experimental observation of extremely low insertion loss linear optical isolation.}
					\textbf{a.} Probe power transmission coefficient $\left| \tilde{t}_p \right|^2 $ is measured in the forward direction through the waveguide near the ($\omega_2$, k$_2$) mode, with fixed 66 $\mu W$ pump power dropped into the ($\omega_1$, k$_1$) mode.
					The forward probe power transmission coefficient through the waveguide shows only 1.44 dB insertion loss within the transparency. The phonon mode frequency is 145 MHz.
					\textbf{b.} The ($\omega_2$, -k$_2$) optical mode measured by the backward probe does not exhibit the induced transparency, resulting in conventional absorption of the probe signal by the resonator. 
					\textbf{c.} The optical isolation contrast is evaluated as the difference between forward and backward power transmission coefficients. Here we calculate 14.4 dB peak contrast with a -3 dB bandwidth of 90 kHz. Isolation exists over 470 kHz.}
				\label{data}
			\end{adjustwidth}
		\end{figure}

		We experimentally demonstrate ultra-low loss optical isolation (Fig.~\ref{data}) through simultaneous forward and backward probing of a silica waveguide and microsphere resonator system (resonator diameter $\sim 170$ \um) in the telecom band 1520-1570 nm. While smaller resonators may also be used, or indeed designed, here we selected the diameter to guarantee the natural existence of multiple triplets of acoustic and optical modes that satisfy the phase-matching condition for BSIT. The experiment is performed at room temperature and atmospheric pressure condition.
		{The microsphere resonator is fabricated by reflow of a single-ended optical fiber taper using an arc discharge. Additionally, a tapered optical fiber waveguide of diameter 1 - 3} \um {is fabricated by linear tension drawing of SMF-28 fiber while being heated with a hydrogen flame, till the point that it supports only a single optical mode with significant evanescent field.} 
		{This fiber mode is used to couple optical signals with the resonator by means of evanescent field overlap with the resonator whispering gallery modes. The optical coupling rate is controlled using distance with a piezo-nanopositioner.}
		With adiabatic tapering \cite{134196} the loss associated with this waveguide can be made as low as 0.003~dB~\cite{Kato2015}. 
		To measure the isolation, the probe power transmission coefficient $\left| \tilde{t}_p \right|^2 $ is quantified in both directions while a constant control driving field \sinf{1} is supplied (see Supplement for detailed experimental setup).
		In this experiment the two selected optical modes of the resonator have linewidth $\kappa_1 \approx \kappa_2 \approx 4.1$~MHz, and are spaced approximately $145$ MHz apart. They are coupled by means of a 145 MHz acoustic mode of intrinsic linewidth $\Gamma_{\textrm{B}} \approx 12$ kHz. Through finite element simulations, we estimate that the acoustic mode corresponds to a first order Rayleigh surface acoustic exictation having an azimuthal order of M=24. At a diameter of 170 $\mu m$, this translates to an acoustic momentum of $q = 0.28 ~\textrm{\textmu}m^{-1}$ and ensures breaking of interaction symmetry for co-propagating and counter-propagating probe fields (Fig.~\ref{wk_3lvl}a).
		
		The requisite BSIT phase-matching is first experimentally verified by strongly driving the ($\omega_2$, k$_2$) optical mode and observing spontaneous and stimulated Stokes Brillouin scattering into the lower mode ($\omega_1$, k$_1$) in the forward direction \cite{Bahl:2011cf}. 
		Subsequently, we drive the ($\omega_1$, k$_1$) optical mode with a fixed control laser and use a co-propagating probe laser to measure the power transmission spectrum across the high frequency optical mode ($\omega_2$,~k$_2$) revealing the induced transparency window.
		The control laser detuning and power are adjusted in order to maximize the power transmission within the transparency peak.
		Experimental measurements of the probe power transmission $\left| \tilde{t}_p \right|^2 $ in both forward and backward directions are presented in Fig.~\ref{data}.
		The system exhibits very low forward insertion loss (1.44 dB) at the peak of induced transparency region for 66 $\mu W$ control laser power absorbed to the resonator (power launched in fiber is 680 $\mu W$). This corresponds to an experimentally calculated pump-enhanced Brillouin coupling of $G \approx \kappa_2 /12$. At this point, the acoustic mode has an effective linewidth of 80.4 kHz due to Brillouin cooling~\cite{Bahl:2012jm}. Simultaneous measurement of backward probe power transmission (Fig.~\ref{data}b) shows only the absorption spectrum of the unperturbed ($\omega_2$,~-k$_2$) optical mode, generating a power transmission loss of $\sim$15.8 dB in the waveguide. 
		Subtraction of the forward and backward measurements provides a measure of the optical isolation contrast, which is 14.4~dB here with $\sim$90 kHz full width at half maximum (Fig.~\ref{data}c). 
		
		\vspace{12pt}
		
		Since the forward insertion loss is very low (zero in the ideal theoretical case), the isolation contrast is primarily determined by the proximity of the waveguide-resonator coupling to the critical coupling condition, which if achieved would yield infinite isolation contrast. 
		Achieving critical coupling $\kex = \kappa_{2,\textrm{o}}$ in non-integrated waveguide-microsphere systems is very challenging due to multimode waveguiding in the taper, thermal drifts during the experiment, and vibrational or mechanical stability issues. Previously, up to 26 dB of signal extinction has been experimentally demonstrated in a fiber taper-microsphere system~\cite{Cai2000}. In the future, ideal isolation may be approached if the waveguide and resonator are integrated on-chip, since most mechanical issues can be eliminated and the interacting modes can be designed precisely.
		Alternatively, applications that require high contrast may employ multiple isolators in series with minimal penalty due to the extremely low insertion loss in this system. 
		It is thus appropriate to compare performance of different isolators by referencing the achieved contrast to 1~dB forward loss. The data shown in Fig.~\ref{data} indicates this figure of merit of approximately 10 $\textrm{dB}_{\textrm{isolation}}/\textrm{dB}_{\textrm{loss}}$ (units preserved for clarity, indicating 14.4 dB constrast vs 1.44 dB forward loss).
		
		\begin{figure}[ht]
			\begin{adjustwidth}{-1in}{-1in}
				\centering
				\includegraphics[scale=0.3]{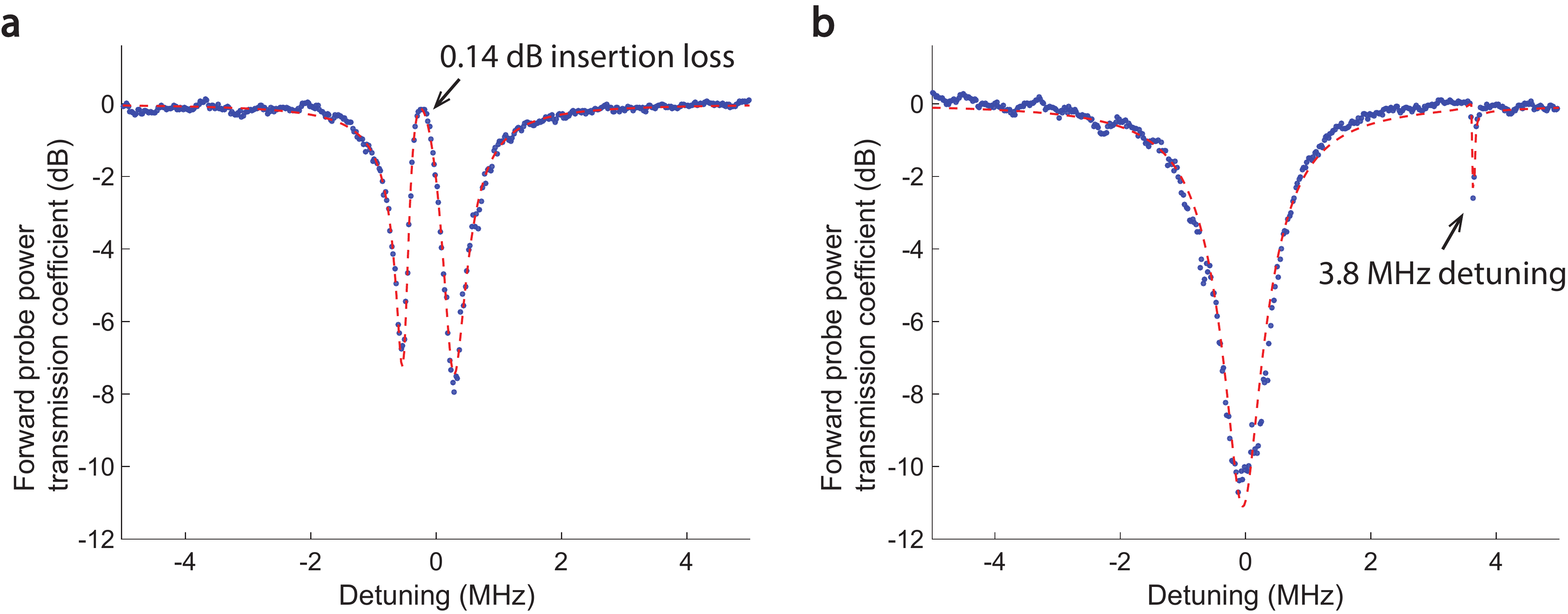}
				\caption{
					\textbf{Demonstration of ultra-low forward insertion loss with stronger coupling $G$.}
					%
					%
					\textbf{a.} 
					Here, we use a triplet of optical and acoustic modes with an optical mode separation or acoustic frequency of 164.8 MHz. Pump-enhanced Brillouin coupling rate $G$ is much higher due to better acousto-optic modal overlap and 235 $\mu W$ power absorbed into the control mode. This results in $G \approx \kappa/3$ causing the forward insertion loss within the transparency to decrease to only 0.14 dB. 
					The isolation bandwidth also increases to approximately 400 kHz. 
					\textbf{b.} The transparency-free ($\omega_2$, k$_2$) optical mode is observable by detuning the control laser from the ($\omega_1$, k$_1$) optical mode, which also detunes the scattered light.}
				\label{data2}
			\end{adjustwidth}
		\end{figure}
		
		\vspace{12pt}
		
		Theory indicates that much lower forward insertion loss can be obtained if {much higher coupling rate $G$ is arranged, either by lowering the loss rates of the optical modes}, or by using higher control laser power.
		{Fortunately, a special feature of two-mode systems such as BSIT} \cite{2016arXiv160707921K} {is the resonant enhancement of the intracavity pump photons in mode $\omega_1$, which enables much easier access to the strong coupling regime. Nonreciprocity based on single-mode OMIT} \cite{Shen2016} {does not possess this feature and it is thus impractical to expand the isolation bandwidth and reach the ultra-low loss regime.}
		{Making use of this resonant enhancement}, in Fig.~\ref{data2}a we show a system nearly reaching the strong coupling regime with $G \approx \kappa /3$, exhibiting only 0.14 dB forward insertion loss (96.8\% transmission) and isolation contrast estimated at 11 dB. Here, 235 $\mu W$ control power is coupled to the resonator (700 $\mu W$ launched in fiber). The unmodified optical mode absorption can be easily observed by detuning the control laser such that the interference is generated outside the optical mode (Fig.~\ref{data2}b). 
		This result indicates that the strong coupling regime is also within the reach of this silica waveguide-resonator system \cite{2016arXiv160707921K}. The isolation figure of merit (referenced to 1 dB insertion loss) for the Fig.~\ref{data2} result is quantified at 78.6 $\textrm{dB}_{\textrm{isolation}}/\textrm{dB}_{\textrm{loss}}$. 
		This compares extremely well to commercial fiber-optic Faraday isolators whose figures of merit typically range between 60 - 100 $\textrm{dB}_{\textrm{isolation}}/\textrm{dB}_{\textrm{loss}}$, and far exceeds the capabilities demonstrated till date by any other non-magnetic microscale optical isolation approach.
		{As shown in Supplementary Table S.1, our achieved contrast exceeds the next best microscale experimental result in non-magnetic optical isolation} \cite{Sayrin:2015bm} {by nearly 7 orders-of-magnitude (69.5 dB difference, i.e. 78.6 dB vs 9.09 dB) per 1 dB of insertion loss.}

		Finally, we also demonstrate in the Supplement the optical reconfigurability of the isolation direction by means of independent control lasers that propagate in opposite directions. 
		{OMIT-based nonreciprocity}~\cite{Shen2016} {does not possess the capability of fully independent reconfiguration since  both forward and reverse optical signals interact with the same zero-momentum vibrational mode. Thus photon conversion can occur through an optomechanical dark mode}~\cite{Dong:2012dr} {shared between forward and reverse pumps, i.e. forward (reverse) sources can modify light propagation in the reverse (forward) direction.}

		\vspace{12pt}
		
		Achieving complete linear optical isolation through optical and opto-mechanical interactions that occur in all media, irrespective of crystallinity or amorphicity, material band structure, magnetic bias, or presence of gain, ensures that the technique could be implemented in {nearly} any photonic foundry process with any optical material. 
		{Example systems that could support this isolation approach are released optomechanical resonators with co-integrated waveguides such as those shown in } \cite{Wiederhecker2009}.
		Since the isolation bandwidth {demonstrated} here is relatively narrow, but is wavelength agnostic, this approach must be tailored for particular photonic device applications.
		{However, we must emphasize} that the maximum bandwidth of this {isolation approach} under strong acousto-optical coupling is only limited by the optical mode linewidth $\kappa_2$, {allowing future improvement in isolation bandwidth to several GHz with the use of low optical Q-factor modes and higher acoustic frequencies}.
		In contrast to all previous works, this induced transparency approach ensures that bidirectional signals are attenuated by default, and only unidirectional transport is enabled when the control optical stimulus is applied. This scheme additionally ensures protection for the system to be isolated in case of failure of the control source, {and allows the possibility of dynamic optical shuttering}.
		The absence of magnetic or radiofrequency electromagnetic driving fields make this approach particularly useful for chip-scale cold atom microsystems technologies, for both isolation and shuttering of optical signals, and laser protection without loss.

	\section*{Acknowledgements}
	Funding for this research was provided through the DARPA Cold Atom Microsystems (CAMS) program, and the Air Force Office for Scientific Research (AFOSR) Young Investigator program.
	
	\vspace{12pt}

	\def\url#1{}


\clearpage

\title{Supplementary information: \\Complete linear optical isolation at the microscale with ultralow loss}

\author{JunHwan Kim$^\dagger$, Seunghwi Kim$^\dagger$, Gaurav Bahl$^\ast$\\
	\\
	\footnotesize{Mechanical Science and Engineering,}\\
	\footnotesize{University of Illinois at Urbana-Champaign, Urbana, Illinois, USA}\\
	\footnotesize{$^\dagger$ Equal contribution.} \\
	\footnotesize{$^\ast$ To whom correspondence should be addressed; E-mail: bahl@illinois.edu.}
}

\date{}
\maketitle

\vspace{12pt} 
	
	\section{Classical description of induced transparency arising from light-sound coupling in a resonator}
	
	The coupled classical field equations for our waveguide-resonator system can be derived from the acoustic and electromagnetic wave equations under the slowly varying amplitude approximation, and a detailed explanation can be found in \cite{PhysRevA.88.013815} and also the Supplement of Ref.~\cite{Kim2015}.
	
	\begin{equation}
	\begin{aligned}
	\dot{a_1} & = 	- (\kappa_1 /2 + j \Delta_1) a_1  -  j \beta^* u^* a_2 + \sqrt{\kex} \, \sinf{1} \\
	\dot{a_2} & = 	- (\kappa_2 /2 + j \Delta_2) a_2  -  j \beta u a_1  + \sqrt{\kex} \, \sinf{2} \\
	\dot{u} & = - (\Gamma_B/2 + j \Delta_B) u	-	j \beta^* a_1^* a_2 + \xi	\\
	\soutf{i} & = \left.	\sinf{i} - \sqrt{\kex} \, a_i ~ \right|_{\textrm{ where } i = 1,2}   		\label{AS_setup1}
	\end{aligned}
	\end{equation}
	
	where $a_i$ is the non-dimensional intracavity optical field at the control ($i$=1) or the probe ($i$=2) frequencies, $u$ is the non-dimensional intracavity acoustic field, $\beta$ is the acousto-optic coupling rate, $\kappa_{i}$ is the loaded optical loss rate, $\Gamma_B$ is the phonon loss rate, and \kex~ is the coupling rate between the waveguide and resonator. The loaded optical loss rates are defined as $\kappa_i = \kappa_{i,\textrm{o}} + \kappa_{ex}$ where $\kappa_{i,\textrm{o}}$ is the loss rate intrinsic to the optical mode. The $\Delta_{i}$ parameters are the field detuning (subscript $B$ for the acoustic field), $\xi$ is the thermal mechanical fluctuation (noise), and \sinf{i} and \soutf{i} are the optical driving and output fields in the waveguide respectively (Manuscript Fig.~1b). 
	We can safely assume that $s_{\textrm{1,in}}$ is a stronger source for the control field $a_1$ compared against the scattering contribution. Further, the thermal fluctuation source $\xi$ is negligible when we solve for time-averaged intracavity fields~\cite{Weis:2010ci}.
	Solving Eqn.~set~\ref{AS_setup1} for the probe output field $s_{\textrm{2,out}}$ given a probe driving field $s_{\textrm{2,in}}$ yields the steady-state probe transmission coefficient $\tilde{t}_p$.
	
	\begin{align}
	\tilde{t}_p ~ = ~ \frac{\soutf{2}}{\sinf{2}} ~ = ~ 1 - \frac{ \kex }{\left( \kappa_2/2 + j \Delta_2 \right) +G^{2}/\left( \Gamma_B/2 + j \Delta_B \right)}
	\label{eq:AS_transmission}
	\end{align}
	
	where $G=|\beta a_1 |$ is the pump-enhanced Brillouin coupling rate. The coupling rate $G$ is manipulated by the control driving field $s_{\textrm{1,in}}$ in the waveguide through the relation $a_1=s_{\textrm{1,in}} \sqrt{\kappa_{\textrm{ex}}} / \left( \kappa_1/2 + j \Delta_1 \right)$. 
	
	The shape of the function presented in Eqn.~\ref{eq:AS_transmission} resembles the conventional optical absorption by a resonator, but with the acousto-optic interaction leading to a transparency within the absorption signature (see Fig.~2 in the Manuscript). As summarized in the Manuscript, the momentum dependence of the acousto-optic coupling rate $\beta$ breaks the direction symmetry of this transparency. When the system is critically coupled and Brillouin acousto-optic coupling is engaged, the transmission $\tilde{t}_p$ can be shown to reach zero in one direction, with transmission in the opposite direction approaching 100\%.
	
	\vspace{12pt}

	\section{Experimental setup}

	We experimentally measure optical isolation produced in the waveguide-resonator system by probing optical transmission through the waveguide in the forward and backward directions simultaneously. 
	We first enable the requisite Brillouin coupling by supplying a relatively strong control laser ($<$1 mW) on the lower frequency optical mode ($\omega_1$, k$_1$) of the system (see manuscript). A co-propagating probe laser then measures the light transmission as it sweeps through the anti-Stokes optical mode ($\omega_2$, k$_2$). When the Brillouin phase-matching condition between the optical and acoustic modes is satisfied, we observe the induced transparency~\cite{Kim2015}. To show optical isolation, the same measurement is taken in the forward and backward directions while the control laser is supplied in the forward direction only.
	
\begin{figure}[h!]
		\vspace{6pt}
		\begin{adjustwidth}{-0.5in}{-0.5in}
			\vspace{-10pt}
			\centering

			\includegraphics[scale=0.221]{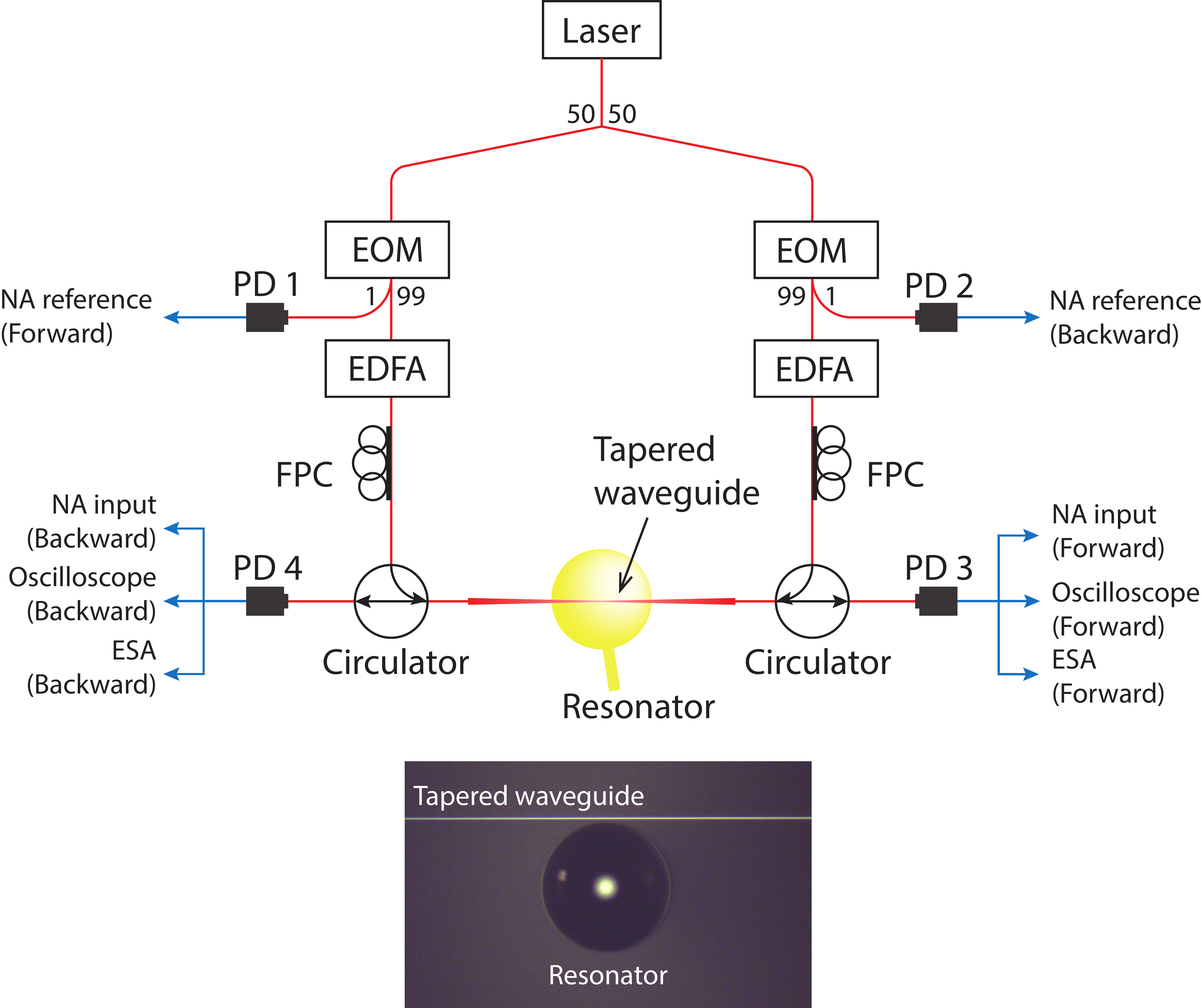}
			\caption{Experimental setup for simultaneous forward and backward probe transmission measurements is shown. We use a matched set of optical components including the electro-optic modulator (EOM), the erbium-doped fiber amplifier (EDFA), the fiber polarization contrller (FPC), the circulator and the photodetectors (PD) for the forward and backward measurements. Light is coupled to the resonator via tapered waveguide. An electronic network analyzer (NA) performs ratiometric measurements of NA reference and NA input signals (marked in figure) in the forward and backward directions. The electrical spectrum analyzer (ESA) is also used to observe the acoustic phonon mode by measuring the beat note generated by the control laser and Brillouin light scattering by the phonons.}
			\label{expSetup}
		\end{adjustwidth}
	\end{figure}

	The experimental setup used for the simultaneous forward and backward measurements is shown in Fig.~\ref{expSetup}. 
	We employ a 1520-1570 nm external cavity diode laser (ECDL) to generate the control and probe laser fields. This laser source is first split into the forward and backward directions using a 50:50 splitter. 
	Electro-optic modulators (EOM) are employed as variable optical attenuators in dc mode (i.e. by adjusting the bias voltage) for manipulating control laser power in either direction. 
	The probe laser is also derived from the control laser using the same EOMs to generate two sidebands spectrally separated from the control by the modulation frequency $\omega_m$. The probe laser frequency $\omega_p = \omega_c + \omega_m$ can be swept using $\omega_m$ relative to the control laser $\omega_c$. 
	An erbium-doped fiber amplifier (EDFA) is used after each EOM to independently modify the control laser power, which in turn regulates the pump-enhanced Brillouin coupling rate $G$ in either direction. 
	Fiber polarization controllers (FPC) are used to match the light polarizations of the forward and backward propagating laser fields. 
	Two circulators are placed before and after the resonator to allow simultaneous measurements of the probe transmissions in the forward and backward directions without reconfiguring the experimental setup. We use a total of four photodetectors, two for measuring the forward and backward probes which are used as references (PD1 and PD2 in Fig.~\ref{expSetup}) and the other two for measuring the forward and backward probe transmissions through the resonator-waveguide system (PD3 and PD4 in Fig.~\ref{expSetup}). The tapered waveguide is fabricated by adiabatic linear drawing of telecom fiber (Corning SMF-28) while heating with a hydrogen flame to a diameter comparable to the laser wavelength~\cite{Ward2014}.
	
	\FloatBarrier
	\vspace{24pt}

	\begin{figure}[b!]
		\begin{adjustwidth}{-0.5in}{-0.5in}
			\centering
			\includegraphics[scale=0.72]{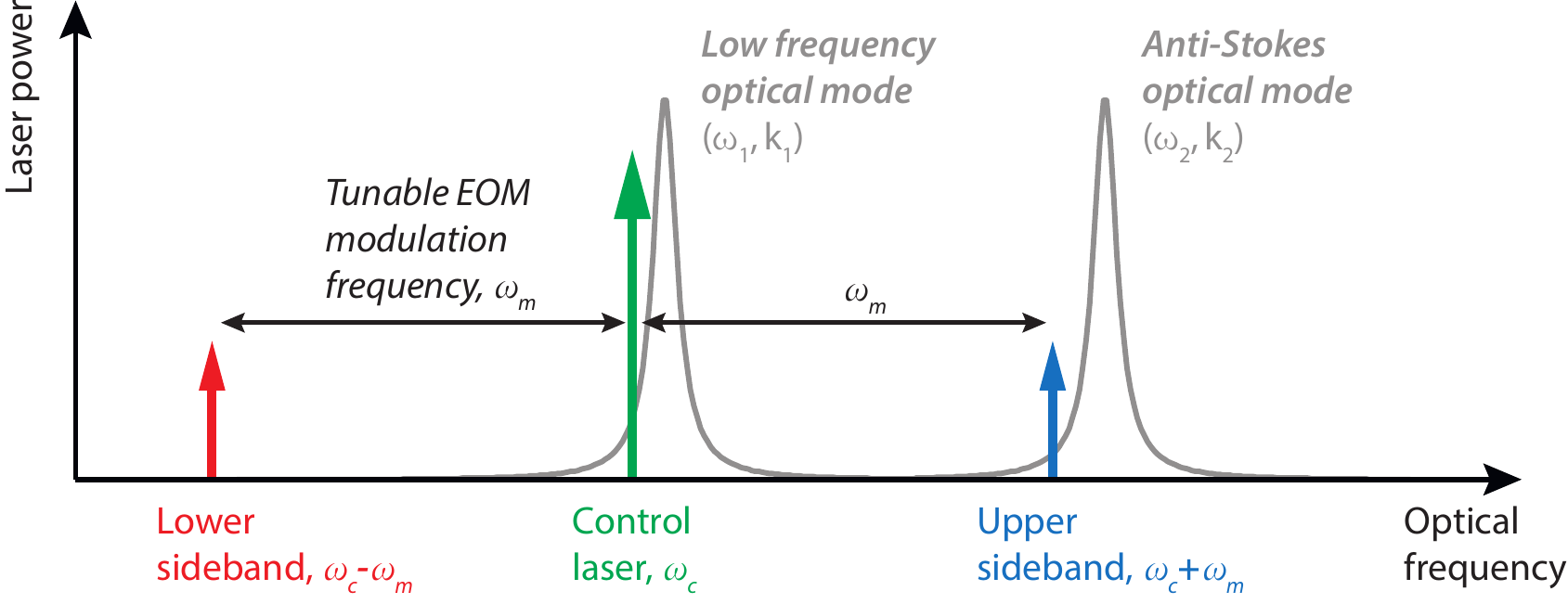}
			\caption{An electro-optic modulator (EOM) generates lower and upper sidebands at modulation offset $\omega_{m}$ away from the control laser. The upper sideband is used as a probe to measure transmission coefficient through the system. The lower sideband does not interact with any feature of interest in this system and transmits unhindered. The optical modes are thermally self-locked~\cite{Carmon2004} to the control laser on the lower optical mode.}
			\label{sidebands}
		\end{adjustwidth}
	\end{figure}

	\section{Determining the optical probe transmission coefficient}
	
	The probe transmission coefficient $\tilde{t}_p$ is measured with the help of a network analyzer, which performs a coherent ratiometric analysis of beat notes of the control and probe optical signals at various points in the experiment (Fig.~\ref{expSetup}). Here, we dissect the network analyzer measurement to explain how the probe transmission coefficient is extracted from the experimental data.

	The control laser with frequency $\omega_c$ is electro-optically modulated at $\omega_m$ to create two sidebands. By keeping the modulation depth low, we can ensure the sidebands are small compared to the carrier, allowing a first-order approximation of the spectrum. We can thus write the optical field within the fiber \textbf{prior to the resonator} as the following
	\begin{align}
	\tilde{E}_{in} & = E_c e^{-j \omega_c t} \left(1 + \frac{m}{2} e^{j \omega_m t} + \frac{m}{2} e^{-j \omega_m t} \right) + c.c.		\label{eqn:Ein}
	\end{align}
	where $E_c$ is the carrier or control laser field amplitude and $m$ is the modulation depth. 
	The control laser frequency $\omega_c$ and modulation frequency $\omega_m$ are adjusted such that the control laser and upper sideband overlap with optical modes while the lower sideband does not couple to any of the resonator's optical modes (Fig.~\ref{sidebands}). We use the upper sideband as the optical probe to measure light transmission through the system, while the control laser parked at the lower frequency optical mode enables the Brillouin scattering interaction between the optical mode pair. 
	
	The optical field arriving at the photodetector \textbf{after passing through the taper-resonator system} (PD3 in the forward case, PD4 in the backward case -- Fig.~\ref{expSetup}) can then be expressed as
	
	\begin{align}
	\tilde{E}_{out} = E_c e^{-j \omega_c t} \left( \tilde{t}_c + \tilde{t}_{ls} \frac{m}{2} e^{j \omega_m t}   + \tilde{t}_{p} \frac{m}{2} e^{-j \omega_m t} \right) + c.c.		\label{eqn:Eout}
	\end{align}
	where $\tilde{t}_i$ are the complex valued transmission coefficients of the control, lower sideband, and probe (upper sideband) fields ($i$=$c$, $ls$, $p$ respectively) through the waveguide. 
	Since the lower sideband does not couple to the resonator, its transmission coefficient is simply $\tilde{t}_{ls} = 1$ (Fig.~\ref{sidebands}). 
	The probe transmission coefficient $\tilde{t}_p$ measured in the forward and backward directions defines the optical isolation performance.

	The optical power measured at the output detector (PD3 or PD4 depending on the probe direction) can be extracted from Eqn.~\ref{eqn:Eout} as shown below. Here we consider only the terms that fall within the detector bandwidth at frequency $\omega_m$.
	
	\begin{align}
	P_{out} \propto \left| { \tilde{E}_{out} } \right|^2 & = \left| E_c \right|^2 t_c \frac{m}{2} \left( e^{-j \omega_m t} + \tilde{t}_p e^{-j \omega_m t} \right) + c.c. \notag \\
	& = \left| E_c \right|^2 t_c m \left[ ~ \left(1 + \operatorname{Re}(\tilde{t}_p)\right) \cos{\omega_m t}  ~+~  \operatorname{Im}(\tilde{t}_p) \sin{\omega_m t} ~  \right]		\label{eqn:outsignal}
	\end{align}
	Without loss of generality we have set $\tilde{t}_c$ with a phase of zero, i.e. all other fields are referenced to the phase of control field. The RF (electrical) output signal from the photodetector is $P_{out}$ multiplied by the detector gain.
	
	\vspace{12pt}
	
	The network analyzer requires a reference signal to perform the ratiometric measurement. We generate this reference by directly measuring the optical signal prior to the resonator, i.e. Eq.~\ref{eqn:Ein}, at PD1 (PD2 in the backward direction, see Fig.~\ref{expSetup} for details). As above, this reference signal is proportional to the optical power
	
	\begin{align}
	P_{ref} ~ \propto ~ 2 \left|E_c\right|^2 m \cos(\omega_m t)
	\end{align}
	With respect to this $\cos(\omega_m t)$ reference, the first term in the parentheses in Eq.~\ref{eqn:outsignal} provides the in-phase component $\left(1+\operatorname{Re}(\tilde{t}_p) \right)$ of the measurement, while the second term provides the quadrature component $\operatorname{Im}(\tilde{t}_p)$. The network analyzer output typically converts this measurement to the polar form
	
	\begin{align}
	A e^{j \phi} = \frac{t_c M}{2} \left( 1 + \tilde{t}_p \right)\label{NAmeasurement}
	\end{align}
	where $A$ is the amplitude response, $\phi$ is the phase response, and $M$ is a coefficient accounting for a fractional difference between optical powers measured at the reference photodetector and at the photodetector placed after the resonator at a decoupled state. $M$ includes EOM output power split ratio (1:99), EDFA gain, and difference in photodetectors' sensitivities. 
	
	The frequency and power of the control laser remain unchanged during the experiment, resulting in a constant $t_c$ that can be determined by monitoring the control laser transmission. $M$ and $t_c$ can also be determined together through the network analyzer response when the probe is off resonance from the anti-Stokes mode, i.e. in the case where $\tilde{t}_{p} = 1$. Using this information, curve fitting can be performed on the network analyzer measurement, and the complex $\tilde{t}_{p}$ can be separately determined as a function of offset (from the control laser) frequency $\omega_m$. However, we note that the transmission coefficient extracted using Eq.~\ref{NAmeasurement} is not of the probe field only. We must also consider the effect of coherent light sources other than the probe for the accurate measurement of the true probe transmission coefficient. We discuss this in detail in the next section.

	\vspace{12pt}

	\section{Background light in probe measurements}
	
	\begin{figure}[b!]
		\begin{adjustwidth}{-0.5in}{-0.5in}
			\centering
			\includegraphics[scale=0.5]{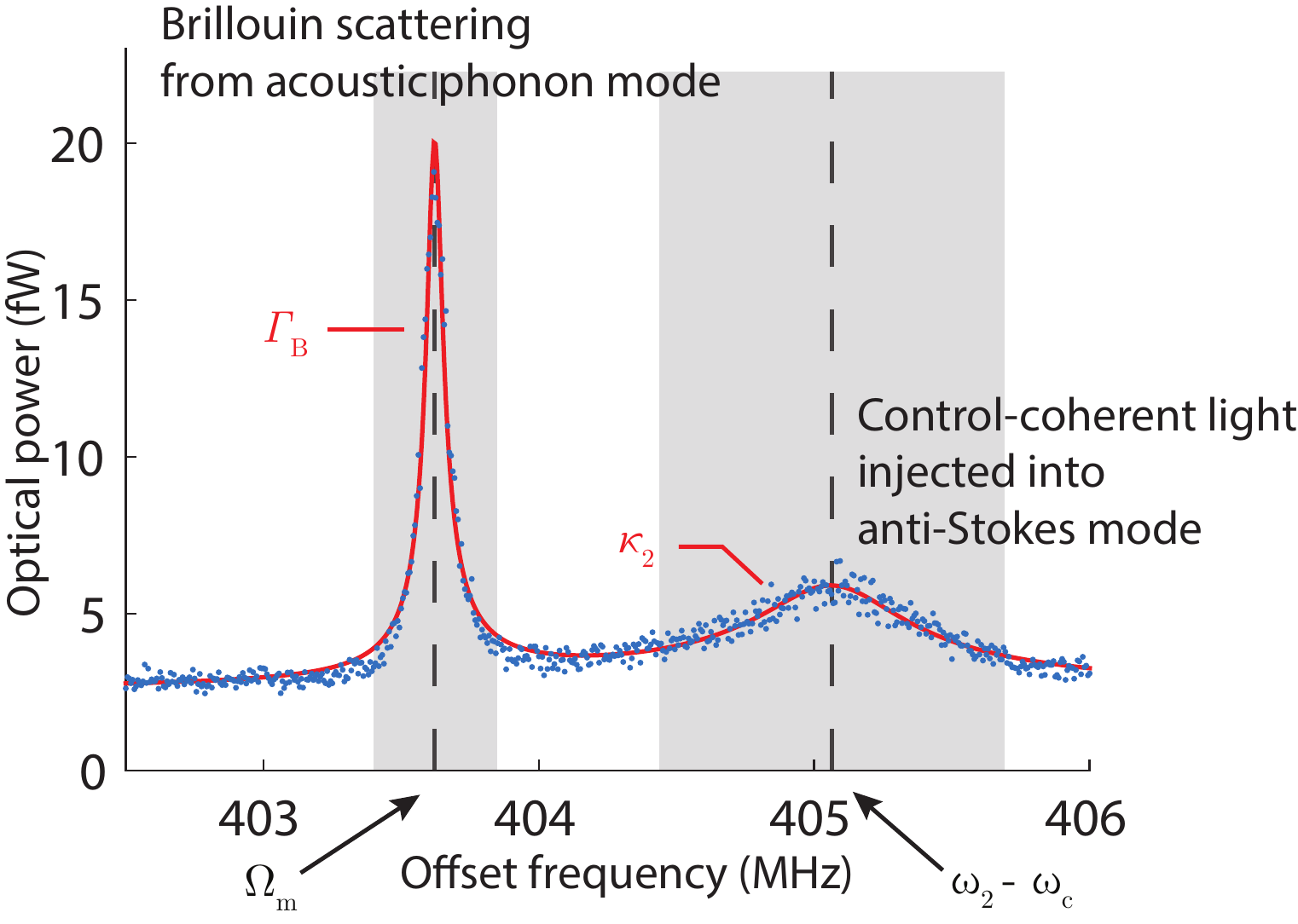}
			\caption{ Measurement of Brillouin scattering at $\Omega_m$ and the control-coherent background light (from defect induced scattering and direct injection) shaped by the anti-Stokes optical mode at $\omega_{2}$. Both sources are offset from the control laser by roughly 404 MHz. This measurement is derived from their beating with the control field on a photodetector and measured by an RF electronic spectrum analyzer. The additional background light is typically too small to be observed except when it is resonantly amplified by an ultra-high-Q resonator. The Brillouin scattering occurs at a fixed frequency defined by the phonon mode $\Omega_m$ while the background light is tunable by modifying the control laser frequency.}
			\label{RSA}
		\end{adjustwidth}
	\end{figure}

\begin{figure}[t]
		\vspace{10pt}
		\begin{adjustwidth}{-0.5in}{-0.5in}
			\centering
			\includegraphics[scale=0.6]{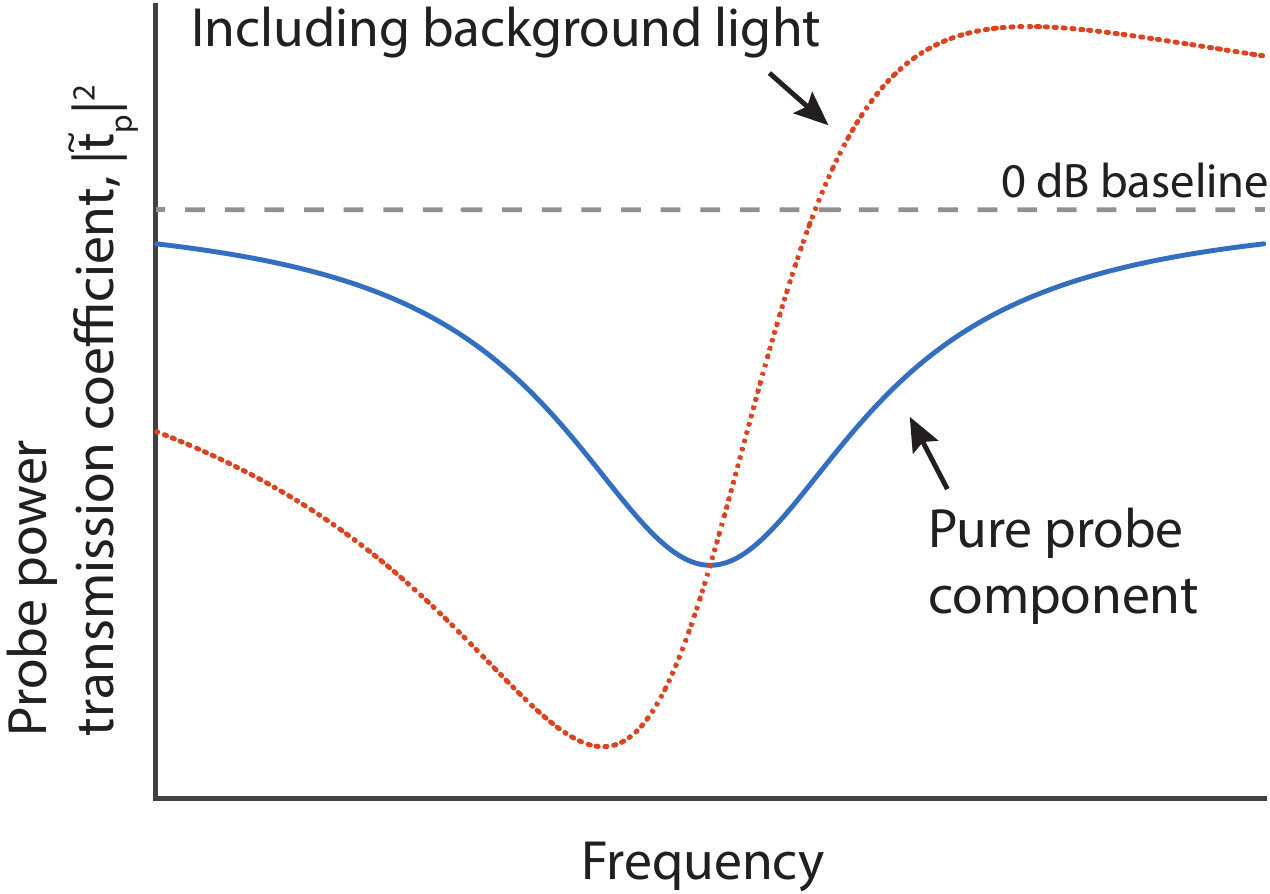}
			\caption{The network analyzer measurement shows an asymmetric probe power transmission coefficient $\left| \tilde{t}_p \right|^2$  (red-dotted line) in spite of the optical mode being symmetric. The distortion in the measurement originates from the background light injected to the anti-Stokes optical mode, creating difficulty in the estimation of the 0 dB transmission baseline (grey-dashed line) and optical mode center frequency. After correcting for the background light, the symmetric optical transmission is seen (blue-solid line). }
			\label{coherentBackground}
		\end{adjustwidth}
	\end{figure}

	In ultra-high-Q resonators, coherent spontaneous light scattering from the control laser by small intrinsic defects can populate the anti-Stokes optical mode. There may also be direct injection of the control laser into the anti-Stokes optical mode. While this extra light is generally small, it does result in competition with the small amount of anti-Stokes light scattering from the acoustic mode in the structure (the phenomenon of interest) and can contaminate measurements of $\tilde{t}_{p}$. An exemplary measurement of this spurious light is shown in Fig.~\ref{RSA}. Since both light sources are being generated from the same pump/control laser, there can be interference that complicates the measurement of the probe transmission (Fig.~\ref{coherentBackground}).

	We can easily observe the evidence of this additional light in the background using an electrical spectrum analyzer that monitors the RF power measurement from the photodetector. Fig.~\ref{RSA} shows the spectrum of beat notes generated from scattering by the acoustic phonon mode centered at $\Omega_m$ = 403.6 MHz, and from background light within the anti-Stokes optical mode centered at $\omega_{as}$ = 405.1 MHz offset from the control laser frequency. Since the resonant frequency of the phonon mode is fixed, the frequency of the beat note originating from Brillouin scattering does not change when the control laser frequency changes. However, the beat note generated from the control-coherent background light injected into the anti-Stokes optical mode can be moved in frequency space with the control laser.

	Since the proposed sources of background light (defect scattering, direct injection) are proportional to the intracavity control field $a_1$, we can model them as a coherent source driving the anti-Stokes mode with coupling strength $r$ relative to the control field. This extra source can be included in the equations describing our system as follows :

	\begin{equation}
	\begin{aligned}
	\dot{a_1} & = 	- (\kappa_1 /2 + j \Delta_1) a_1  -  j \beta^* u^* a_2 + \sqrt{k_{ex}} s_{1,in} \\
	\dot{a_2} & = 	- (\kappa_2 /2 + j \Delta_2) a_2  -  j \beta u a_1  + \sqrt{k_{ex}} s_{2,in} + j r a_{1}\\
	\dot{u} & = - (\Gamma_B/2 + j \Delta_B) u	-	j \beta^* a_1^* a_2 + \xi	\\
	s_{i,out} & = s_{i,in} - \sqrt{k_{ex}} \, a_i ~ |_{\textrm{ where } i = 1,2}
	\label{AS_setup}
	\end{aligned}
	\end{equation}
	
	As discussed in the Supplement \S 1, the control field scattering $j \beta^* u^* a_2$ and thermal fluctuation $\xi$ can be assumed to be negligible. We find the modified steady-state intracavity probe field $a_2$ which is composed of the unperturbed probe response and an additional background light term:
	
	\begin{align}
	a_2 & = \frac{\sqrt{k_{ex}} s_{2,in}}{\gamma_{2} + G^2/\gamma_{B}} + \frac{j r  a_{1} }{\gamma_{2} + G^2/\gamma_{B}}\label{AS_a_2_asym}\\
	\gamma_i & = \kappa_i/2 +j\Delta_i		\notag \\
	\gamma_B & = \Gamma_B/2 + j\Delta_B 		\notag
	\end{align}
	where $G = \left|\beta a_1\right|$ is the pump-enhanced Brillouin coupling rate. The probe field arriving at the photodetector is then expressed as
	
	\begin{align}
	s_{2,out} & = \left(1 - \frac{k_{ex}}{\gamma_2 + G^2/\gamma_B}\right) s_{2,in}  -  \left(\frac{j r \sqrt{k_{ex}}}{ \gamma_2 + G^2/\gamma_B } \right) a_{1}	\label{sout_asymmetry}  \\
	& =  \tilde{t}_{p,\textrm{actual}} \, s_{2,in} ~ - ~  \left(\frac{j r \sqrt{k_{ex}}}{ \gamma_2 + G^2/\gamma_B } \right) a_{1}	\notag
	\end{align}
	implying that the measured probe transmission coefficient (by definition) will be
	
	\begin{align}
	\tilde{t}_{p, \textrm{measured}} ~ & = ~ \frac{s_{2,out}}{s_{2,in}} \notag \\
	& = ~  \tilde{t}_{p,\textrm{actual}} ~ - ~  \left(\frac{j r \sqrt{k_{ex}}}{ \gamma_2 + G^2/\gamma_B } \right) \frac{a_{1}}{s_{2,in}}\label{measured_tp}
	\end{align}
	From Eq.~\ref{measured_tp}, we note that we can reduce the effect of the background light by increasing the probe laser power (larger $s_{2,in}$) during the experiment such that the first term dominates. 
	The measured transmission coefficient $\tilde{t}_{p,\textrm{measured}}$ acquired from Eq.~\ref{NAmeasurement} will then approach the actual transmission coefficient $\tilde{t}_{p,\textrm{actual}}$. More generally, however, the second interfering term results in an asymmetry (irrespective of acousto-optical coupling) in the optical transmission through high-Q resonators measured by this pump-probe technique.

	\begin{figure}[t]
		\begin{adjustwidth}{-0.5in}{-0.5in}
			\centering
			\includegraphics[width=1.2\textwidth]{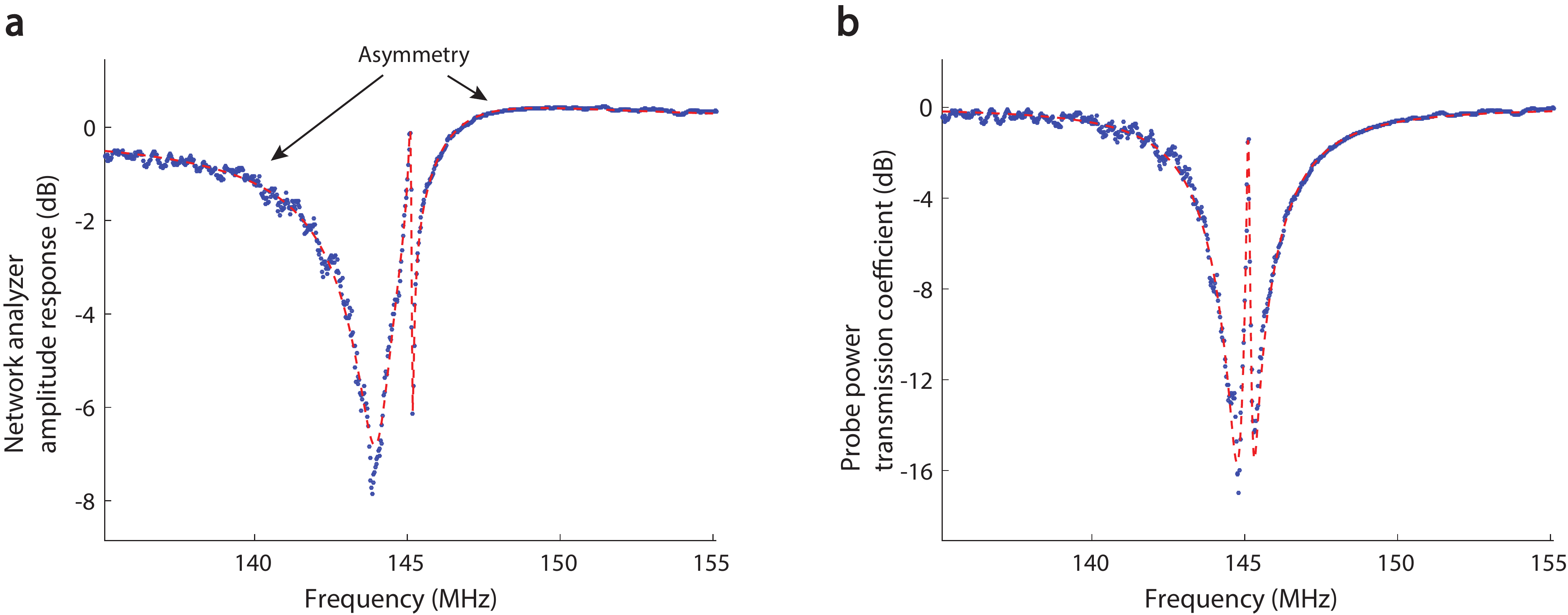}
			\caption{
				\textbf{a.} Raw amplitude response data from the network analyzer. The optical transmission measurement and the induced transparency are distorted by the additional light within the anti-Stokes optical mode.
				\textbf{b.} Probe optical power transmission is plotted using input-output relation after removal of the background light. Data shows transparency within the Lorentzian shaped optical mode. The red dashed line represents a fit using theoretical model for induced transparency.}
			\label{coherentBackgroundData}
		\end{adjustwidth}
	\end{figure}
	
	In Fig.~\ref{coherentBackground} we plot Eq.~\ref{sout_asymmetry} with acousto-optical coupling set to zero ($G=0$). The extra light in the resonator modifies the transmission coefficient such that the high frequency side of the optical mode exceeds 0~dB baseline while the low frequency side is artificially dipped. When we exclude the background light, the plot now reveals a symmetric optical mode and a true resonance frequency. Thus, factoring out this asymmetry is critical in accurately determining the optical isolation performance. 
	Unprocessed transmission measurement from the network analyzer (Fig.~\ref{coherentBackgroundData}a) shows the asymmetric optical mode shape and different 0~dB baseline levels on either side of the resonance. Such a mismatch in baseline is used to estimate the degree of asymmetry and the coupling strength $r$. The background light can then be subtracted from the measurement to obtain $\tilde{t}_{p,actual}$ as shown in Fig.~\ref{coherentBackgroundData}b. In the Fig.~\ref{coherentBackgroundData} example, a symmetric optical mode with transparency at the center of the optical mode is revealed.
	
	\vspace{24pt}
	

	\section{Demonstrating reconfiguration of optical isolation}
	
\begin{figure}[h]
		\begin{adjustwidth}{-0.5in}{-0.5in}
			\centering
			\includegraphics[width=1.2\textwidth]{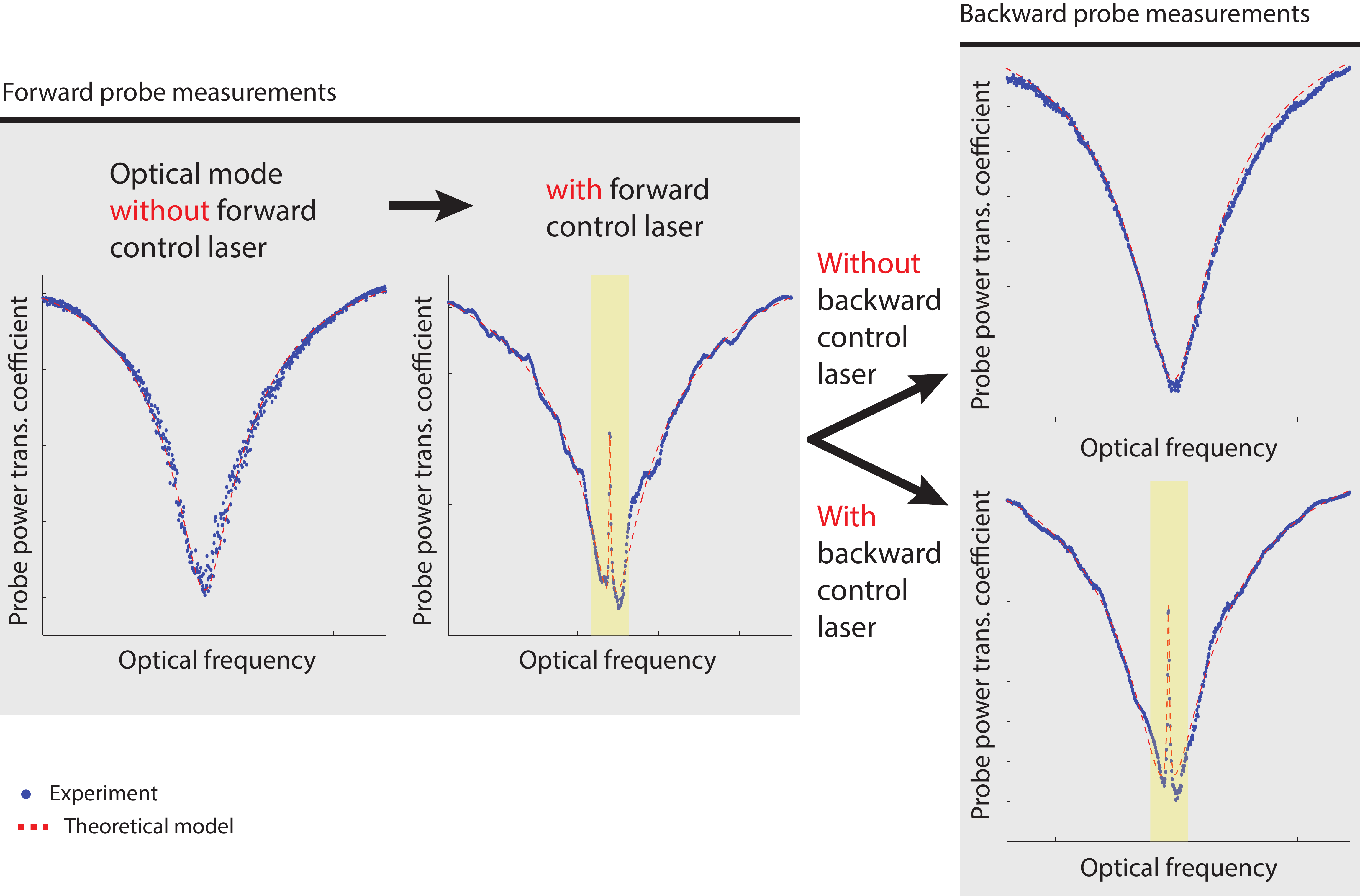}
			\caption{Demonstrating reconfigurable optical isolation. Increasing the control laser power in the forward direction, we observe the appearance of the acousto-optical transparency. While transparency is enabled in the forward direction, we can switch on and off the transparency in the backward direction using a separate backward propagating control laser. The red dashed line represents a fit using theoretical model for induced transparency.}
			\label{reconfig}
		\end{adjustwidth}
	\end{figure}

	The directionality of optical isolation can be modified in either direction by choosing an appropriate control laser. This is simply demonstrated through an experiment (Fig.~\ref{reconfig}) where the the control laser field is sequentially provided in the forward direction only, backward direction only, and in both directions simultaneously. Since the forward and backward directions in a whispering-gallery resonator are nominally decoupled and the phonon mode also has an associated directionality (i.e. momentum), the transparency is independently observed in the directions in which a control laser field is supplied.

	Fig.~\ref{reconfig} shows that when no control field is provided, the anti-Stokes optical mode is a simple Lorentzian shaped dip. However, when the control laser is supplied in the forward direction, a transparency is observed by the forward probe. While this transparency is sustained in the forward direction, we can independently switch on and off the transparency in the backward direction. This is demonstrated by probing the anti-Stokes optical mode in the backward direction with and without a backward control laser, which results in an optical mode with and without transparency respectively. Such reconfigurable transparency has never previously been demonstrated in any other optical or opto-mechanical system.

	\FloatBarrier

	\vspace{20pt}
	
	\newpage
	
	\section{Comparison of isolation performance}
	
	We quantitatively compare isolation performance against previously demonstrated linear, magnet-free optical isolators in Supplementary Table S.1. 
	Isolation contrast (extinction ratio), quantifies the ratiometric difference for the forward and reverse transmitted optical signals. 
	Insertion loss quantifies the difference in the input and forward transmitted signals.
	All these approaches only operate over finite bandwidth, for which the 3 dB bandwidth quantifies the frequency span over which the contrast is within 3 dB of its highest value. 
	We also provide device size and the system used to assist with determining the fit for specific applications.
	Since isolators may also be cascaded to increase contrast, we normalize the isolation contrast for each demonstration using 1 dB of insertion loss as a reference point, and provide the figure of merit as dB of contrast per 1 dB of insertion loss.

	While many previous reports show signatures of optical nonreciprocity, several do not quantify the contrast or insertion loss metrics making it difficult to have a direct comparison. 
	Sayrin et. al. report an excellent contrast per 1 dB of insertion loss from a resonantly enhanced spin-polarized cold atom system~\cite{Sayrin:2015bm}. However, the target applications are different from the other magnet-free isolation approaches reported here as this requires laser cooling of the system. 
	Lira et. al. report an impressive bandwidth of 200 GHz, but only show 3 dB contrast and extremely high insertion loss~\cite{Lira:2012ck}.
	
	Our result exhibits an enormous 78.6 dB contrast per 1 dB of insertion loss, which rivals the values seen on commercial magneto-optic based optical isolators. Our demonstration has 69.5 dB higher figure of merit (nearly 7 orders-of-magnitude) relative to the next best microscale isolator result \cite{Sayrin:2015bm}.

	\begin{landscape}
		\begin{centering}
			\begin{table}[ht!]
				
				\vspace{-50pt}
				\caption{Comparison of isolation performance for experimentally demonstrated non-magnetic linear non-reciprocal systems.}

				\begin{tabular}{| >{\raggedright\arraybackslash}m{3cm} | >{\raggedright\arraybackslash}m{2.4cm} | >{\raggedright\arraybackslash}m{1.7cm} | >{\raggedright\arraybackslash}m{1.9cm} | >{\raggedright\arraybackslash}m{1.6cm} | >{\raggedright\arraybackslash}m{2.2cm} | >{\raggedright\arraybackslash}m{2.2cm} |
						>{\raggedright\arraybackslash}m{2.2cm} |}
					
					\hline
					Author & Technique & Isolation contrast & Insertion loss & 3 dB Bandwidth & Device size & Contrast (dB) per 1 dB insertion loss (dB/dB) & System used \\ \hline \hline
					\multicolumn{8}{|l|}{ Isolation using photonic microdevice } \\ \hlineB{2.5}
					This work & BSIT & 11 dB & 0.14 dB & 400 kHz & 170 \um & 78.6 & Microsphere \\ \hline
					H. Lira et. al.~\cite{Lira:2012ck} & Interband scattering & 3 dB & 70 dB & 200 GHz & 110 \um & 0.0429 & Waveguide on-chip \\ \hline
					J. Kim  et. al.~\cite{Kim2015} & BSIT & \cellcolor{black!15} N/A & \cellcolor{black!15} N/A & 16.9 kHz & 150 \um & \cellcolor{black!15} N/A & Microsphere \\ \hline
					C.-H. Dong et. al.~\cite{Dong2015} & BSIT & \cellcolor{black!15} N/A & \cellcolor{black!15} N/A & 9 kHz & 196 \um & \cellcolor{black!15} N/A & Microsphere \\ \hline
					Z. Shen et. al.\cite{Shen2016} & OMIT & \cellcolor{black!15} N/A & \cellcolor{black!15} N/A & 22 kHz & 36 \um & \cellcolor{black!15} N/A & Microsphere \\ \hline
					B. Peng et. al.~\cite{Peng2014} & PT symmetry breaking & \cellcolor{black!15} N/A & \cellcolor{black!15} N/A & \cellcolor{black!15} N/A & 60 \um$^\dagger$ & \cellcolor{black!15} N/A & Toroid on-chip \\ \hline
					L. D. Tzuang et. al.~\cite{Tzuang2014} & Photonic Aharonov-Bohm effect & 2.4 dB & \cellcolor{black!15} N/A & \cellcolor{black!15} N/A & 325 \um & \cellcolor{black!15} N/A & Waveguide on-chip \\ \hline \hline
					
					\multicolumn{8}{|l|}{ Isolation through atom interactions } \\ \hlineB{2.5}
					C. Sayrin et. al.~\cite{Sayrin:2015bm} & Scattering from spin-polarized cold atoms & 7.8 dB & 1.08 dB & \cellcolor{black!15} N/A & 0$^*$ & 7.22 & Atoms on tapered fiber \\ \hline
					C. Sayrin et. al.~\cite{Sayrin:2015bm} & Same as above with resonant enhancement & 13 dB & 1.43 dB & \cellcolor{black!15} N/A & 36 \um & 9.09 & Atoms on resonator \\ \hline \hline
					
					\multicolumn{8}{|l|}{ Isolation in macroscale fiber system } \\ \hlineB{2.5}
					M. S. Kang et. al.~\cite{Kang:2011el} & Stimulated Brillouin scattering & 20 dB & \cellcolor{black!15} N/A  & 7 MHz & 15 m & \cellcolor{black!15} N/A & Photonic crystal fiber \\ \hline

				\end{tabular}
				
				\vspace{12pt}
				$^\dagger$ Requires two resonators of size 60 um.
				
				$^*$ Size of an atom.
				
				N/A = Not available.

				\label{FOM_table}
				
			\end{table}
		\end{centering}
	\end{landscape}

\vspace{24pt}

\FloatBarrier

\end{document}